\documentclass[conference]{IEEEtran}

\usepackage{myStyleIEEE}

\IEEEoverridecommandlockouts

\makeatletter

\begin{document}

\title{Linear State Estimation and Bad Data Detection for Power Systems with RTU and PMU Measurements}
\author{
\IEEEauthorblockN{Aleksandar Jovicic, Gabriela Hug}%
%\IEEEauthorblockA{\IEEEauthorrefmark{2} Faculty of Electrical Engineering, Technion-Israel Institute of Technology, Haifa 3200003, Israel}
\IEEEauthorblockA{EEH - Power Systems Laboratory, ETH Zurich, Zurich, Switzerland } %\\
\{jovicic, hug\}@eeh.ee.ethz.ch\vspace*{-0.525cm}}

\maketitle

\begin{abstract}
In this paper, a novel linear algorithm is proposed for state estimation including bad data detection of power systems that are monitored both by conventional and synchrophasor measurements. Both types of data are treated simultaneously and the states are estimated in rectangular coordinates. The proposed estimator is based on the linear weighted least square method. To enable the derivation of linear measurement functions, the network is modelled in terms of voltages and currents in rectangular form and pseudo-measurements are used to represent conventional measurements. Furthermore, the largest normalized residual test is used to detect bad data. To validate the accuracy and robustness of the proposed algorithm, several test cases of different sizes are solved and the results are presented and discussed.    

\end{abstract}


\iffalse
% Use this to place sponsorships
\thanksto{\noindent Submitted to the 21st Power Systems Computation Conference (PSCC 2020).}
\fi

\section{Introduction}
State Estimation (SE) is a mathematical algorithm that has a crucial role in power system monitoring. SE processes raw measurement data collected from measurement devices installed throughout the grid and provides an estimate of the system states, i.e. voltage magnitudes and angles for all system buses. Based on these estimates, the system operator gains insight into the actual operating state of the grid. 

The measurement data have been obtained for decades by Remote Terminal Units (RTU), which measure voltage magnitudes, as well as active and reactive power injections and line flows. To estimate the state of the system, these types of data are most commonly processed by a Weighted Least Square (WLS) algorithm, as initially proposed in \cite{Schweppe}. WLS minimizes the weighted mismatch between the measured values and their corresponding measurement functions that relate them to the system states. These functions are based on power flow equations, and are therefore highly nonlinear. 

The SE area has been a matter of great research interest lately, due to an increased utilization of Phasor Measurement Units (PMU) in transmission networks, which diversified the available measurement set. PMUs provide highly accurate measurements of current and voltage phasors. Additionally, the measured values are synchronised in time via Global Positioning System, which further enhances the accuracy. Therefore, achieving full system observability solely by PMUs would be an ideal scenario, since the states could be estimated with a high precision. Furthermore, this would render the SE problem linear \cite{Phadke1985}, which drastically improves the computational efficiency. However, this scenario is unlikely to happen in the foreseeable future, mainly due to the high cost of a PMU installation \cite{DOE}. Consequently, novel hybrid SE methods are needed that will be able to include measurements provided both by RTUs and PMUs, thereby leveraging the broad system coverage by the former, and increased accuracy of the latter.  

Many hybrid estimators were proposed in literature to address the aforementioned problem. The multi-stage estimators in \cite{Ming2006,Baltensperger2010,Costa2013} process conventional and synchrophasor measurements separately. In \cite{Ming2006,Baltensperger2010} this is done sequentially, while the algorithm in \cite{Costa2013} utilizes two separate estimators for RTU and PMU measurements that are executed in parallel, which is followed by a fusion stage to produce final estimates. Another group of hybrid SE methods are single-stage estimators that account for all types of measurements simultaneously. Different approaches for PMU current measurement transformation are used in \cite{Chakrabarti2010,Asprou2011}, while the algorithm in \cite{Valverde2011} incorporates all measurements directly, which is enabled by the expansion of the state vector to include current magnitudes and angles. All these approaches are based on the nonlinear WLS algorithm, and therefore need to employ iterative solution methods.

Recent advances in the field of power system simulation include the efforts to apply concepts of circuit theory to a range of power system related problems \cite{CMU1,CMU2}. By leveraging this idea, novel hybrid circuit-based estimators were derived in \cite{Jovicic,Jovicic2,Wagner2019}. These are characterized by simultaneous treatment of conventional and synchrophasor measurements, and by states being estimated in rectangular coordinates. Furthermore, the methods in \cite{Jovicic2,Wagner2019} propose fully linear estimation algorithms, resulting in a substantially decreased computational time. However, these approaches are not robust against bad data.

Raw measurement data contain random errors due to imperfections of the measuring equipment and disturbances in the communication channels. Filtering these errors is the primary task of any state estimator. However, the capability of the SE algorithm to detect gross measurement errors and suppress their negative effect on the estimation accuracy is also an important feature. All previously listed algorithms, including the circuit-based estimators, are not bad data resilient. In general, this is a well-known drawback of all estimators based on the WLS approach. To overcome this issue, a post-processing step called the Largest Normalized Residual (LNR) test is most commonly utilized to detect and identify bad data \cite{Abur}. This approach was used in \cite{Korres2011} for the nonlinear WLS-based estimator which treats both RTU and PMU data. The main drawback of the LNR test is that it detects only one bad measurement per iteration. Therefore, the estimation process has to be repeated until all bad data have been detected, which, in combination with the inherent nonlinearity of the WLS-based estimators, results in a significant computational burden.   

To avoid the use of the post-processing step, a separate class of estimators that are robust against outliers are proposed in \cite{Bad1,Bad2,Bad5}. These are based on the Least Absolute Value algorithm, which in general has a very high computational complexity, and is vulnerable to leverage points \cite{Abur}. Alternatively, the robust method proposed in \cite{Bad4} employs separate Maximum Correntropy Criterion-based estimators for RTU and PMU data, followed by a fusion stage to generate final estimates. An iterative solution approach is still needed.

In this paper, a novel linear state estimation algorithm including bad data detection is derived for systems that comprise both conventional and synchrophasor measurements. Both types of measured data are processed simultaneously, and the states are estimated in rectangular coordinates. The proposed estimator utilizes a linear WLS framework, which yields very high computational speed and scalability of the algorithm. This is enabled by representing RTU data by pseudo-measurements that are linearly related to the system states. Finally, LNR testing is employed in the post-processing stage to suppress the negative effect of bad data that might occur in the measurement set.

The remainder of the paper is structured as follows. Section \ref{sec: 2} gives an overview of the conventional WLS method, along with the description of the LNR test for bad data detection. Furthermore, the approach for linear modelling of RTU data used in circuit-based estimators is explained. The proposed linear SE algorithm is presented in Sect. \ref{sec: 3}. Section \ref{sec: 4} showcases simulation results, and the main conclusions are summarized in Sect. \ref{sec: 5}.

\section{Technical Background} \label{sec: 2}
\subsection{Conventional WLS-based state estimation} \label{sec: 2a}
Power system state estimation is traditionally formulated as a Weighted Least Square problem. The vector of states that are estimated comprises angles ($\boldsymbol{\theta}$) and magnitudes ($\boldsymbol{V}$) of the bus voltages:
\begin{equation} \label{eq: state_vector}
    \boldsymbol{x} = \begin{bmatrix}
    \boldsymbol{\theta} \\ \boldsymbol{V}
    \end{bmatrix}
\end{equation}
The voltage angle of an arbitrarily chosen bus is usually set to zero to serve as a reference, and is therefore removed from the state vector.

A measurement set $\boldsymbol{z}$ is represented as:
\begin{equation} \label{eq: basic_model}
    \boldsymbol{z} = \boldsymbol{h(x)} + \boldsymbol{e} 
\end{equation}
where $\boldsymbol{h(x)}$ are measurement functions relating measurement values to the state variables, and $\boldsymbol{e}$ is the vector of measurement errors, which are assumed to be normally distributed and uncorrelated. \iffalse Measurement functions $\boldsymbol{h(x)}$ for the measurements of injection powers and power flows are derived from the power flow equations and are therefore nonlinear. On the other hand, in the case of voltage and current phasor measurement, the corresponding functions are linear. \fi

By minimizing the weighted sum of squares of the measurement residuals, i.e. by solving the following optimization problem:
\begin{equation} \label{eq: WLS_obj}
    \text{min}\,\,\boldsymbol{J(x)} = \left[\boldsymbol{z} - \boldsymbol{h(x)}\right]^T\boldsymbol{R}^{-1}\left[\boldsymbol{z} - \boldsymbol{h(x)}\right]
\end{equation}
the most probable state of the system can be found. Therein, $\boldsymbol{R}$ is a diagonal measurement error covariance matrix and for each measurement $i$, $R_{i,i}$ is equal to the square of the standard deviation $\sigma$ of the corresponding measurement. Hence, $\boldsymbol{R}^{-1}$ comprises weight coefficients for all measurements.

In order to find the optimal solution, i.e. the optimal state vector, the first-order optimality conditions are derived for \eqref{eq: WLS_obj}. Measurement functions $\boldsymbol{h(x)}$ for the measurements of injected powers and line flows are derived from the power flow equations and are therefore nonlinear, thus rendering the entire optimization problem nonlinear. As a result, an iterative procedure has to be utilized to solve it. The following equation holds at each iteration $\textit{k}$: 
\begin{equation} \label{eq: normal}
    \left[\boldsymbol{G}(\boldsymbol{x}^k)\right]\Delta\boldsymbol{x}^{k} = \boldsymbol{H}^T(\boldsymbol{x}^k)\boldsymbol{R}^{-1}\left[\boldsymbol{z} - \boldsymbol{h}(\boldsymbol{x}^k)\right]
\end{equation}
where $\boldsymbol{H}(\boldsymbol{x}^k)=\partial\boldsymbol{h}(\boldsymbol{x}^k)/\partial\boldsymbol{x}^k$, $\Delta\boldsymbol{x}^{k}=\boldsymbol{x}^{k+1}-\boldsymbol{x}^{k}$ and $\boldsymbol{G}(\boldsymbol{x}^k)=\boldsymbol{H}^T(\boldsymbol{x}^k)\boldsymbol{R}^{-1}\boldsymbol{H}(\boldsymbol{x}^k)$. Equation \eqref{eq: normal} is solved for $\Delta\boldsymbol{x}^{k}$ at each iteration, until its value is smaller than a predefined convergence threshold.

Alternatively, if the WLS problem is linear, e.g. if the measurement set comprises only PMU measurements of voltage and current phasors and the state vector is formulated in rectangular coordinates \cite{Phadke1985}, there is no need to execute an iterative procedure, as the measurement functions $\boldsymbol{h(x)}$ are linear and the system states can be estimated directly by solving the following equation for $\boldsymbol{x}$:
\begin{equation} \label{eq: linearWLS}
    \boldsymbol{x} = \left(\boldsymbol{\boldsymbol{H}^T\boldsymbol{R}^{-1}\boldsymbol{H}}\right)^{-1}\boldsymbol{H}^T\boldsymbol{R}^{-1}\boldsymbol{z}
\end{equation}
where $\boldsymbol{H}$ is a constant Jacobian matrix.

\subsection{Bad data detection and identification} \label{sec: 2_1}
As mentioned, certain measurements may contain large errors, which can occur due to various reasons. Therefore, one of the main tasks of any state estimator is to detect the existence of outliers in the measurement set and suppress their negative effect on the accuracy of the state estimates. 

A common method to detect bad data in WLS-based SE algorithms is using normalized residuals \cite{Abur}. The first step is to solve the SE problem and calculate the measurement residuals $r_i$ for each measurement $i$:
\begin{equation} \label{eq: residual}
    r_i = z_i - h_i(\hat{\boldsymbol{x}})
\end{equation}
where $\hat{\boldsymbol{x}}$ is the vector of estimated states. Then, normalized residuals are calculated as:
\begin{equation} \label{eq: norm_residual}
    r^N_i = \frac{|r_i|}{\sqrt{\Omega_{i,i}}} \iffalse\quad\quad\quad     \forall i \in \{1,...,m\} \fi
\end{equation}
where $\boldsymbol{\Omega}$ is the residual covariance matrix, computed as $\boldsymbol{\Omega}=\boldsymbol{R}-\boldsymbol{H}\left(\boldsymbol{H}^T\boldsymbol{R}^{-1}\boldsymbol{H}\right)^{-1}\boldsymbol{H}^T$. The so obtained normalized residual vector $\boldsymbol{r^N}$ has a standard normal distribution, since it is assumed that measurement errors have Gaussian distribution and are uncorrelated. Thus, the existence of bad data in the measurement set can be detected by finding the largest normalized residual, $r^N_{\text{max}}$, and comparing its value against a predefined threshold $q$. If $r^N_{\text{max}} < q$, no bad data is detected and the estimation process is completed. However, if $r^N_{\text{max}} > q$, the corresponding measurement is identified as bad data and is removed from the measurement set. The SE algorithm is then executed again until no bad data is detected.   

Alternatively, if the measurement $b$ is identified as bad data, instead of removing it from the measurement set, it can be corrected by subtracting the estimated value of its error from the actual measured value $z_b$. The corrected value of the measurement $b$ is therefore calculated as: 
\begin{equation} \label{eq: correct_res}
    z^{\text{cor}}_b = z_b - \frac{R_{b,b}}{\Omega_{b,b}}r_b 
\end{equation} 
The SE procedure is repeated after the correction is made. For more details, the reader is referred to \cite{Abur}.

\subsection{Linear modelling of RTU measurements} \label{sec: linearmodel}
An Equivalent Circuit Formulation (ECF) for the power flow problem has been recently proposed in \cite{CMU1,CMU2}. The main idea of this method is that an entire power system can be represented by an equivalent circuit in terms of voltage and current state variables in rectangular coordinates. As a result, the equivalent circuit consists of two coupled sub-circuits, where the first sub-circuit represents real voltages and currents, while the second one comprises their imaginary parts. The ECF approach was applied to the state estimation problem in \cite{Jovicic} for the first time. The resulting nonlinear estimator treated RTU and PMU measurements simultaneously and provided state estimates in rectangular coordinates. The only nonlinearity in this formulation was associated with the circuit representation of RTU measurements. This work was leveraged in \cite{Jovicic2} to derive a fully linear estimator by reformulating the circuit model for RTU measurements. A similar approach was also proposed in \cite{Wagner2019}. In this paper, while still relying on the circuit approach for RTU modelling, we formulate the measurement functions such that \eqref{eq: linearWLS} becomes directly applicable.  

Directly deriving measurement functions for the power measurements of RTUs yields nonlinear functions in terms of real and imaginary voltages and currents. Hence, if an RTU bus $k$ is observed and the measurement set comprises bus voltage magnitude $(V_k)$, and active $(P_{k})$ and reactive $(Q_{k})$ power injections, the following relations between injected current and voltage at this bus are derived \cite{Jovicic2}:
\begin{align}
    I_{R,k} &= \frac{P_{k}}{V_k^2}V_{R,k} + \frac{Q_{k}}{V_k^2}V_{I,k} \label{eq: I_R}\\
    I_{I,k} &= \frac{P_{k}}{V_k^2}V_{I,k} - \frac{Q_{k}}{V_k^2}V_{R,k} \label{eq: I_I}
\end{align}
where $V_{R,k}$ and $V_{I,k}$ are real and imaginary voltages at bus $k$, and $I_{R,k}$ and $I_{I,k}$ are real and imaginary injected currents\footnote{The expressions \eqref{eq: I_R}-\eqref{eq: I_I} were derived for the case where the reference directions of measured voltage and current correspond to load conditions.}. Since $V_k$, $P_k$ and $Q_k$ are measurement values, these equations are linear in terms of the system states.

As an additional preparatory step, we introduce two equations based on Kirchhoff's current law (KCL), one for real and one for imaginary currents, for each RTU bus $k$:
\begin{align}
    I_{R,k} + \sum_{i\in N_{inc,k}}I_{R,i} + I_{R,err,k} = 0\label{eq: KCL_R}\\
    I_{I,k} + \sum_{i\in N_{inc,k}}I_{I,i} + I_{I,err,k} = 0\label{eq: KCL_I}
\end{align}
where $I_{R,i}$ and $I_{I,i}$ are real and imaginary currents flowing through the network components $i$, i.e. shunts, transmission lines and transformers, that are incident to the observed RTU bus $k$, and $I_{R,err,k}$ and $I_{I,err,k}$ represent error terms that need to be minimized. Hence, in an ideal case where all measurements are perfectly accurate, $I_{R,err,k}$ and $I_{I,err,k}$ are equal to zero. Each current in \eqref{eq: KCL_R}-\eqref{eq: KCL_I} has a positive sign if its reference direction is out of the bus, and negative if it is directed into the bus. In case line flows are available instead of injected powers, the same modelling technique can be used. Finally, it is important to mention that the previously described modelling method assumes that the voltage magnitude is available for each RTU bus, which is a reasonable assumption since voltage has to be measured in order to obtain power data. For more details on linear RTU modelling used in the circuit-based estimators, the reader is referred to \cite{Jovicic2,Wagner2019}.      

\section{Proposed Linear State Estimator} \label{sec: 3}
To achieve a linear SE formulation, the state vector in the proposed algorithm consists of bus voltages in rectangular coordinates: 
\begin{equation} \label{eq: hybrid_state_vector}
    \boldsymbol{x} = \begin{bmatrix}
    \boldsymbol{V_R} \\ \boldsymbol{V_I}
    \end{bmatrix}
\end{equation}
where $\boldsymbol{V_R}$ and $\boldsymbol{V_I}$ are vectors of real and imaginary bus voltages. For a system with $N$ buses, the size of vector $\boldsymbol{V_R}$ is $N$, while there are $N-1$ elements in $\boldsymbol{V_I}$, since one bus serves as the reference for all angles, and its imaginary voltage is therefore fixed. The proposed SE method utilizes linear WLS, and therefore obtains the estimate of the system states by solving \eqref{eq: linearWLS} for $\boldsymbol{x}$. The structure of individual terms in \eqref{eq: linearWLS} used in the proposed method will be discussed below.

\subsection{Measurement vector $\boldsymbol{z}$} \label{sec: 3_A}
Conventional and synchrophasor measurements are treated simultaneously, and the vector of measurements which is processed by the proposed method is:
\begin{equation} \label{eq: measurements_org}
    \boldsymbol{z_{org}} = \begin{bmatrix}
    \boldsymbol{z_{PMU}} \\ \boldsymbol{z_{RTU}} 
    \end{bmatrix}
\end{equation}
where $\boldsymbol{z_{PMU}}$ is the vector of PMU measurements, consisting of voltages $(\boldsymbol{V_{R,PMU}}$, $\boldsymbol{V_{I,PMU}})$ and currents ($\boldsymbol{I_{R,PMU}}$, $\boldsymbol{I_{I,PMU}}$) in rectangular coordinates, while $\boldsymbol{z_{RTU}}$ is the vector of RTU measurements comprising voltage magnitudes ($\boldsymbol{V}$), as well as active and reactive injections $(\boldsymbol{P_{inj}}$, $\boldsymbol{Q_{inj}})$ and line flows $(\boldsymbol{P_{flow}}$, $\boldsymbol{Q_{flow}})$. It is important to emphasize that the vector of originally available measurements $\boldsymbol{z_{org}}$ is not the same as $\boldsymbol{z}$ as we use in \eqref{eq: linearWLS}. While both comprise raw PMU measured values, conventional measurements are accounted for in $\boldsymbol{z}$ by representing the original RTU data from $\boldsymbol{z_{org}}$ as pseudo-measurements:
\begin{equation} \label{eq: measurements}
    \boldsymbol{z} = \begin{bmatrix}
    \boldsymbol{z_{PMU}} \\ \boldsymbol{z_{RTU,pseudo}} 
    \end{bmatrix}
\end{equation}
Namely, for each RTU bus, each available group of measurements comprising: (i) bus voltage magnitude, active and reactive power injections; or (ii) bus voltage magnitude, active and reactive line flows in any line incident to the respective bus, is represented in $\boldsymbol{z_{RTU,pseudo}}$ with two pseudo-measurements that are equal to zero. It is assumed that a voltage magnitude measurement is available for each RTU bus. To clarify the idea, an RTU bus with three lines incident to it is observed, and the available measurement set consists of voltage magnitude, as well as active and reactive injections and line flows in all lines. These measurements are represented in $\boldsymbol{z_{RTU,pseudo}}$ with eight pseudo-measurements equal to zero. This lays the foundation for the derivation of linear measurement functions that relate the RTU pseudo-measurements to states, based on the idea discussed in Sect. \ref{sec: linearmodel}. The derivation of these functions will be discussed in the following section.

\subsection{Measurement functions $\boldsymbol{h(x)}$ and Jacobian $\boldsymbol{H}$}
Since the states are estimated in rectangular coordinates, the appropriate models must be derived for all power system components in terms of voltages and currents in rectangular form.

\begin{figure}[t!]
\vspace{-0.7cm}
\centering
\includegraphics[width=0.3\textwidth]{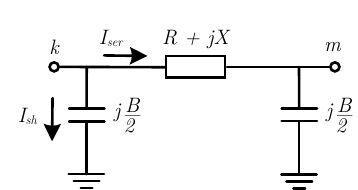}
\caption{Pi-model of a transmission line}
\label{fig:transmission_line}
\vspace{-0.5cm}
\end{figure}

A pi-model of a transmission line connecting buses $k$ and $m$ is given in Fig.~\ref{fig:transmission_line}. To find the relation between real and imaginary voltages and currents of the line, we start with the following general equation:
\begin{equation} \label{eq: Ohm}
\begin{split}
    \underline{I}=\underline{Y}\underline{V}&=\left(Y_R+jY_I\right)\left(V_R+jV_I\right) \\&=Y_RV_R-Y_IV_I+j\left(Y_RV_I+Y_IV_R\right)
\end{split}
\end{equation}
The complex admittance of the series branch is $\underline{Y}_{branch}=\frac{1}{R+jX}=\frac{R}{R^2+X^2}-j\frac{X}{R^2+X^2}$, while the admittance of the shunt branch is $\underline{Y}_{sh}=j\frac{B}{2}$. After incorporating these real and imaginary admittance terms into \eqref{eq: Ohm}, the following expressions for real and imaginary series and shunt branch currents are obtained:
\begin{align}
    I_{R,ser} &= \frac{R}{R^2+X^2}V_{R,ser} + \frac{X}{R^2+X^2}V_{I,ser} \label{eq: I_R_branch}\\
    I_{I,ser} &= \frac{R}{R^2+X^2}V_{I,ser} - \frac{X}{R^2+X^2}V_{R,ser} \label{eq: I_I_branch}\\
    I_{R,sh} &= -\frac{B}{2}V_{I,sh} \label{eq: I_R_shunt}\\
    I_{I,sh} &= \frac{B}{2}V_{R,sh} \label{eq: I_I_shunt}
\end{align}
where $V_{C,ser}$, $C\in\{R,I\}$, is the voltage across the series branch, equal to the difference of voltages at buses $k$ and $m$, while $V_{C,sh}$ is the voltage at the bus where the corresponding shunt is located. One can observe that the currents in \eqref{eq: I_R_branch}-\eqref{eq: I_I_shunt} are linear functions of voltages, since all network parameters are constants. The linear models for other network components, i.e. transformers, phase-shifters, tap-changers and shunt elements, can be derived by applying the same technique, and will not be presented here due to space limitations, but the reader is referred to \cite{CMU1}.  

Once the entire network is modelled in terms of voltages and currents in rectangular form, linear measurement functions $\boldsymbol{h(x)}$ can be defined with respect to the measurement vector $\boldsymbol{z}$ in \eqref{eq: measurements} as follows:

\subsubsection{PMU voltage}
It is assumed in this paper that PMU current and voltage measurements are available in rectangular coordinates. Also, the state vector comprises bus voltages in the same form. Hence, the measurement functions for real and imaginary PMU voltages at bus $k$ are trivial:
\begin{align}
    h^{PMU}_{V_{R,k}} &= V_{R,k}\\
    h^{PMU}_{V_{I,k}} &= V_{I,k}
\end{align}
\subsubsection{PMU injected current}
The measurement functions for the PMU real and imaginary injection currents at bus $k$ are defined according to the KCL:
\begin{align}
    h^{PMU}_{I_{R,inj,k}} &= \sum_{i\in N_{inc,k}}I_{R,i}\\
    h^{PMU}_{I_{I,inj,k}} &= \sum_{i\in N_{inc,k}}I_{I,i}
\end{align}
where $N_{inc,k}$ is the set of all network components incident to bus $k$ and $I_{C,i}$, $C\in\{R,I\}$, are the currents flowing through these components. As mentioned above, these currents are linear functions of the system states. Each current $i$ has a positive sign if it flows out of the bus, while it is negative if it is directed into the bus. The opposite applies to the sign of the measured value in vector $\boldsymbol{z}$. Thus, if the measured injection current flows into the bus, it has a positive sign.
\subsubsection{PMU line flow}
In case a PMU located at bus $k$ measures real and imaginary currents injected into the line connecting this bus to bus $m$, the corresponding measurement functions are defined as:  
\begin{align}
    h^{PMU}_{I_{R,line,km}} &= I_{R,ser,km} + I_{R,sh,k}\\
    h^{PMU}_{I_{I,line,km}} &= I_{I,ser,km} + I_{I,sh,k}
\end{align}
where $I_{C,ser,km}$, $C\in\{R,I\}$, is the current flowing through the series branch from bus $k$ to $m$, and $I_{C,sh,k}$ is the current flowing through the shunt branch incident to bus $k$. The same above-mentioned sign convention still holds.

\subsubsection{RTU power injections} \label{sec: power_inj}
As it was explained in Sect. \ref{sec: 3_A}, if the measurements of voltage magnitude $(V_k)$ and active $(P_{k})$ and reactive $(Q_{k})$ power injections are available at bus $k$, these data are represented in the measurement vector $\boldsymbol{z}$ by two pseudo-measurements equal to zero. These are related to the states based on the approach presented in Sect. \ref{sec: linearmodel}, thus the measurement functions are derived based on the KCL equations for real and imaginary currents at bus $k$. One can observe the analogy between \eqref{eq: basic_model} and \eqref{eq: KCL_R}-\eqref{eq: KCL_I}. Zero values on the right-hand side in \eqref{eq: KCL_R}-\eqref{eq: KCL_I} correspond to the RTU pseudo-measurements in $\boldsymbol{z}$, hence one pseudo-measurement is related to the KCL equation for the real currents, and the other corresponds to the KCL equation for imaginary currents. Mismatch currents $I_{C,err,k}$, $C\in\{R,I\}$, correlate to the error terms $\boldsymbol{e}$ in \eqref{eq: basic_model}, while $I_{C,k} + \sum_{i\in N_{inc,k}}I_{C,i}$ terms essentially represent the measurement functions $\boldsymbol{h(x)}$. Finally, if $I_{R,k}$ and $I_{I,k}$ are replaced with the expressions in \eqref{eq: I_R}-\eqref{eq: I_I}, the following relations are obtained for the measurement functions relating real and imaginary RTU injection pseudo-measurements at bus $k$ to the system states:
\begin{align}
    h^{RTU}_{R,inj,k} &= \frac{P_{k}}{V_k^2}V_{R,k} + \frac{Q_{k}}{V_k^2}V_{I,k}+\sum_{i\in N_{inc,k}}I_{R,i} \label{eq: h_R_RTU_inj}\\
    h^{RTU}_{I,inj,k} &= \frac{P_{k}}{V_k^2}V_{I,k} - \frac{Q_{k}}{V_k^2}V_{R,k} + \sum_{i\in N_{inc,k}}I_{I,i} \label{eq: h_I_RTU_inj}
\end{align}
All currents in expressions \eqref{eq: h_R_RTU_inj}-\eqref{eq: h_I_RTU_inj} have a positive sign if they are directed away from the bus. If the value of $P_{k}$ denotes generation conditions, the term $P_{k}/V_k^2$ will be negative. The same applies to the reactive power and the sign of $Q_{k}/V_k^2$.   

\subsubsection{RTU line flows}
If an RTU located at bus $k$ measures voltage magnitude ($V_k$), as well as active ($P_{km}$) and reactive ($Q_{km}$) line flows in the transmission line connecting this bus to bus $m$, this measurement set will be represented in the measurement vector $\boldsymbol{z}$ by two pseudo-measurements equal to zero. These are related to the system states by applying the same approach as in the case of RTU power injections. Hence, the corresponding measurement functions are:
\begin{equation}
\begin{split}
    h^{RTU}_{R,line,km} = \frac{P_{km}}{V_k^2}V_{R,k} + \frac{Q_{km}}{V_k^2}V_{I,k}\\+I_{R,ser,km} + I_{R,sh,k}\label{eq: h_R_RTU_line}
\end{split}
\end{equation}
\begin{equation}
\begin{split}
    h^{RTU}_{I,line,km} = \frac{P_{km}}{V_k^2}V_{I,k} - \frac{Q_{km}}{V_k^2}V_{R,k}\\+I_{I,ser,km} + I_{I,sh,k}  \label{eq: h_I_RTU_line}
\end{split}
\end{equation}
The sign convention is the same as in the case of power injections. 

A very important feature of the proposed estimator is that all measurement functions are linear, which is achieved by estimating voltages in rectangular coordinates and transforming RTU measurements to yield pseudo-measurements that are linearly related to the states. Finally, the measurement Jacobian is calculated as $\boldsymbol{H}=\partial\boldsymbol{h(x)}/\partial\boldsymbol{x}$, where $\boldsymbol{h(x)}$ is the vector of measurement functions, while $\boldsymbol{x}$ is the vector of states. Since all measurement functions are linear, $\boldsymbol{H}$ is a constant matrix. The size of $\boldsymbol{H}$ is $M\times (2N-1)$, where $M$ is the number of measurements in $\boldsymbol{z}$, and $N$ is the number of buses. 
  
\subsection{Measurement error covariance matrix $\boldsymbol{R}$}
Same as in conventional WLS, the measurement errors are assumed to be normally distributed and uncorrelated. Thus, $\boldsymbol{R}$ is a diagonal matrix in the proposed algorithm, where for each measurement $i$ the corresponding diagonal entry $R_{i,i}$ denotes the variance of the measurement. The size of $\boldsymbol{R}$ is $M\times M$, where $M$ is the number of measurements in $\boldsymbol{z}$. 

For each PMU measurement with standard deviation $\sigma$, the corresponding variance is simply calculated as $\sigma^2$. On the contrary, variances of the RTU pseudo-measurements are calculated based on their measurement functions and the error propagation theory \cite{Navidi}. If bus $k$ is observed and the measurement set consists of voltage magnitude ($V_k$) and active ($P_{k}$) and reactive ($Q_{k}$) power injections, variances of the terms $P_{k}/V_k^2$ and $Q_{k}/V_k^2$ can be calculated based on the following general rule:
\begin{equation}
    f = AB \, \text{or} \, f = \frac{A}{B} \implies \sigma_{f}^2 = f^2\left[\left(\frac{\sigma_{A}}{A}\right)^2+\left(\frac{\sigma_{B}}{B}\right)^2\right] \label{eq: var_product} 
\end{equation}
where the measurements $A$ and $B$ are transformed to the pseudo-measurement $f$, and $\sigma_A$, $\sigma_B$ and $\sigma_f$ are standard deviations of $A$, $B$ and $f$, respectively. Let $\sigma^2_{P,k}$ be the obtained variance of the term $P_{k}/V_k^2$, and let $\sigma^2_{Q,k}$ denote the calculated variance of $Q_{k}/V_k^2$. Then, variances of the real and imaginary RTU pseudo-measurements that represent the observed measurement set are approximated based on their measurement functions \eqref{eq: h_R_RTU_inj}-\eqref{eq: h_I_RTU_inj}. According to the error propagation theory, the properly calculated variance of the real RTU pseudo-measurement would be equal to $V^2_{R,k}\sigma^2_{P,k}+V^2_{I,k}\sigma^2_{Q,k}$, while the variance of the imaginary pseudo-measurement would be equal to $V^2_{I,k}\sigma^2_{P,k}+V^2_{R,k}\sigma^2_{Q,k}$. However, values of rectangular bus voltages are states and are not known prior to the execution of the algorithm. Therefore, the variances of the real and the imaginary pseudo-measurements are set to $\sigma_{P,k}^2$ and $\sigma_{Q,k}^2$, respectively. These values are chosen since, in general, real voltages are substantially higher then their imaginary counterparts, and therefore the corresponding terms in the measurement functions predominantly define the appropriate weight coefficient, i.e. the variance. In case the measurement set comprises measurements of voltage magnitude and active and reactive line flows, variances of the corresponding real and imaginary pseudo-measurements are determined by applying the same approach. 

\subsection{Bad data processing}
Gross errors in either RTU or PMU measurements can severely bias the estimated states. Therefore, it is very important to detect them and suppress their negative effect on the estimation outcome. To this aim, the proposed algorithm utilizes the largest normalized residual test that was explained in Sect. \ref{sec: 2_1}. Normalized measurement residuals are calculated according to \eqref{eq: norm_residual} and the value of the largest normalized residual $r^N_{\text{max}}$ is compared to the threshold $q$. Since the normalized measurement residuals have standard normal distribution, the value of the threshold $q$ is set to 3 \cite{Abur}. If $r^N_{\text{max}}>q$, the corresponding measurement is identified as bad data. Instead of eliminating it from the measurement set, its value is corrected based on \eqref{eq: correct_res} and the estimation process is executed again. This process has to be repeated as long as bad data are detected in the measurement set. However, the overall computational time is still very low due to the linearity of the estimator. The benefit of correcting bad data instead of removing it is that the data structure remains the same. Also, it prevents any RTU pseudo-measurement, which essentially represents a set of three RTU measurements, from being eliminated from the estimation process if any of the original measured values that it is related to has a large error. Therefore, the proposed algorithm is robust against bad data in both RTU and PMU measurement sets. 

\section{Numerical Results} \label{sec: 4}
The performance of the proposed algorithm is evaluated by using the IEEE 14, 57 and 118 bus test systems, and the 2869 and 13659 test systems provided by the PEGASE project \cite{13k_buses}. Accuracy and computational time are examined for the cases where: i) the measurement set consists only of measurements with random Gaussian errors and there are no bad data; ii) bad data are present in the measurement set. The full algorithm is executed in both cases, i.e. both state estimation and bad data detection and correction stages are employed.

\subsection{Test cases without bad data}
The measurement set comprises both conventional and synchrophasor measurements for all test systems. The set of RTU measurements consists of bus voltage magnitudes, as well as active and reactive power injections and line flows. It is assumed that each PMU device has a sufficient number of channels to monitor all lines incident to the bus where it is located. Therefore, the set of PMU measurements comprises phasors of bus voltages and line currents in rectangular coordinates. It should be emphasized that PMU measurements in polar form can also be incorporated, by transforming them into rectangular coordinates and calculating the corresponding variances based on the error propagation theory\cite{Navidi}. In each test case, one PMU bus is selected to serve as the slack bus, thus providing the reference for the voltage angles. The structure of the measurement sets for all test systems is presented in Table \ref{tab:measures}. Obviously, the majority of measurements are provided by RTUs, which corresponds to the current situation in real transmission networks, which are still predominantly monitored by legacy measurements. Locations and number of measurements were selected so that the full system observability is ensured for all test systems.
\begin{table}[t!]
\vspace{-0.4cm}
\centering
\caption{Measurement Sets for Test Cases without Bad Data}
\label{tab:measures}
\begin{tabular}{ |c|c|c|c|c|c| }
 \hline
 {\textbf{Test Case}} & \multicolumn{2}{|c|}{\textbf{PMU}} & \multicolumn{3}{c|}{\textbf{RTU}}\\
 \cline{2-6}
  &\textbf{Voltage} & \textbf{Current} &\textbf{Voltage}  & \textbf{Injection} & \textbf{Flow}\\ 
 \hline
 14 buses & 5 & 14 & 11 & 10 & 36\\
 \hline
 57 buses & 13 & 40 & 47 & 50 & 112\\ 
 \hline
 118 buses & 19 & 76 & 106 & 96 & 298\\ 
 \hline
 2869 buses & 409 & 1362 & 2652 & 2596 & 5134\\ 
 \hline
 13659 buses & 1557 & 5294 & 12870 & 12786 & 25682\\ 
 \hline
\end{tabular}
\vspace{-0.3cm}
\end{table}

All measurement errors are assumed to have Gaussian distribution with zero mean and are not correlated. For each measurement $i$, the measured value is randomly selected from the range $\left[z^t_{i}-\sigma_{i}, z^t_{i}+\sigma_{i}\right]$, where $z^t_{i}$ is the true value of the measurement $i$ obtained from the power flow simulation in MATPOWER, while $\sigma_{i}$ is its corresponding standard deviation. The standard deviations for different types of measurements are given in Table \ref{tab:deviations}.
\begin{table}[t!]
\centering
\caption{Measurement Standard Deviations}
\label{tab:deviations}
\begin{tabular}{|c|c|c|c|c|}
\hline
\multicolumn{2}{|c|}{\textbf{PMU}} & \multicolumn{3}{c|}{\textbf{RTU}}\\
\hline
\textbf{Voltage}   & \textbf{Current}   & \textbf{Voltage}  & \textbf{Injection}   & \textbf{Flow} \\
 \hline
0.02\%  & 0.02\% & 0.4\%   &  1\% & 1\% \\
\hline
\end{tabular}
\vspace{-0.5cm}
\end{table}

Two different performance indices are used in order to evaluate the accuracy of the proposed method. The first one is the sum of variances of the estimated states:
\begin{equation}
    \sigma^2_{x} = \sum^{2N}_{i = 1}(\hat{x}_i - x^t_{i})^2
\end{equation}
where $\hat{x}$ and $x^t$ are estimated and true values, respectively. The second index essentially compares variances of the estimated measurement values and their original values:   
\begin{equation}
    \xi = \frac{\sum^{M}_{i = 1}(\hat{z}_i - z^t_{i})^2}{\sum^{M}_{i = 1}(z^m_{i} - z^t_{i})^2} \label{eq: meas_err}
\end{equation}
where $\hat{z}$, $z^t$ and $z^m$ are the estimated, true and original measurement values, respectively. $M$ denotes the number of measurements. Hence, $\xi<1$ indicates that the estimated values are closer to the true values, compared to the raw measured data. The proposed algorithm is implemented in Matlab, and simulations are executed on a PC with an Intel i7-6600 CPU and 16 GB of RAM. One hundred simulations are performed for each test case to obtain averaged results for different measurement values. The measurement allocation is kept the same for all simulations, while the measurement values are randomly selected as it was explained above. 

\begin{figure}[t!]
\centering
\includegraphics[width=0.5\textwidth]{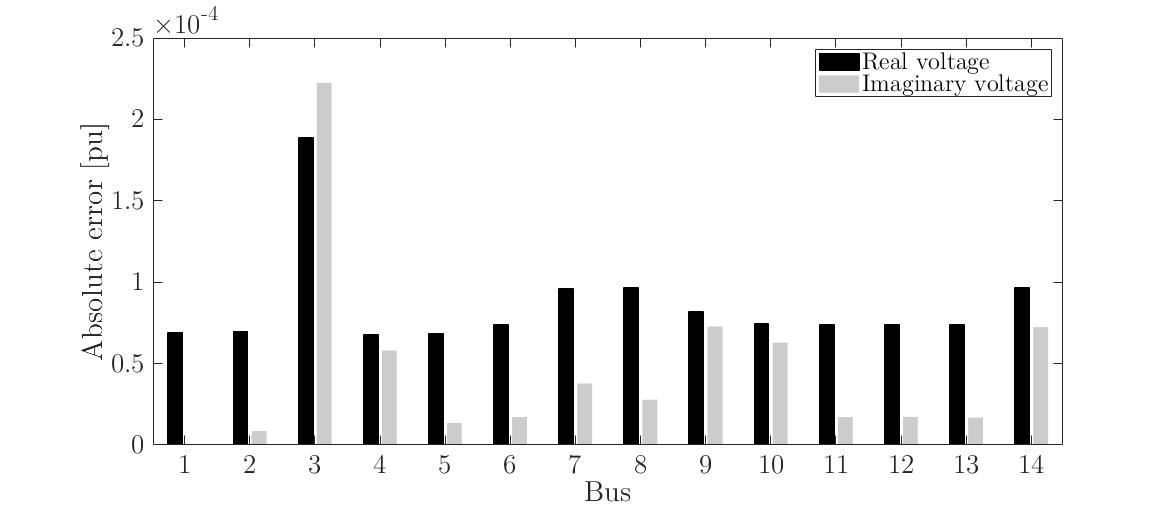}
\caption{State estimation errors for the 14 bus test case}
\label{fig:no_bad}
\vspace{-0.5cm}
\end{figure}

The average values of all performance indices for the cases without bad data are presented in Table \ref{tab:results}. The obtained results demonstrate high accuracy of the proposed method, since the values of all indices are very low for the test cases of all sizes. The values of index $\xi$ show that the estimated measurement values are much more accurate than the original ones. The absolute errors of the estimated bus voltages, both real and imaginary, are presented in Fig. \ref{fig:no_bad} for the 14 bus test case. One can observe very high accuracy of each individual estimated voltage value. Furthermore, the average computational time is given for each test case in Table \ref{tab:results}. The obtained values are very low, which is the consequence of utilization of the linear WLS framework. Also, it is shown that the algorithm is fast even for very large systems, which demonstrates the scalability of the proposed method.

\begin{table}[t]
\vspace{-0.4cm}
\centering
\caption{Average Performance Indices}
\label{tab:results}
\begin{tabular}{ |c|c|c|c| }
 \hline
 \textbf{Test Case} & $\boldsymbol{\sigma^2_x}$ & $\boldsymbol{\xi}$ & \textbf{Time [s]}\\ 
 \hline
 14 buses & 2.7915\,x\,$10^{-7}$ & 0.1183 & 0.0003 \\
 \hline
 57 buses & 2.3162\,x\,$10^{-6}$ & 0.2728 & 0.0015 \\ 
 \hline
 118 buses & 8.1891\,x\,$10^{-6}$ & 0.3248 & 0.0093 \\ 
 \hline
 2869 buses & 1.2373\,x\,$10^{-3}$ & 0.4697 & 0.1065 \\ 
 \hline
 13659 buses & 0.0165 & 0.5827 & 0.5521 \\ 
 \hline
\end{tabular}
\vspace{-0.3cm}
\end{table}

\subsection{Test cases with bad data}
Two different test cases are examined in order to evaluate the capability of the proposed algorithm to accurately estimate the state of the system when bad data are present in the measurement set. Both test cases are based on the IEEE 14 bus test system. Case 1 represents the situation in which there is only one measurement with large error. For this experiment, the same measurement set is used as for the 14 bus test system in the previous section. However, the value of one PMU measurement, namely the real voltage at bus 1, is altered by 30\%. Case 2 considers the presence of multiple bad data. Again, the measurement set for the 14 bus test system from the previous section is used as the initial point. Then, the following PMU and RTU measurements are manually changed: real voltage at bus 1, real line current in the line connecting buses 6 and 5, voltage magnitude at bus 12, active power injection at bus 5, and active and reactive line flows in the line connecting buses 7 and 8. All values are altered by 30\%. 

\begin{figure}[t!]
\centering
\includegraphics[width=0.5\textwidth]{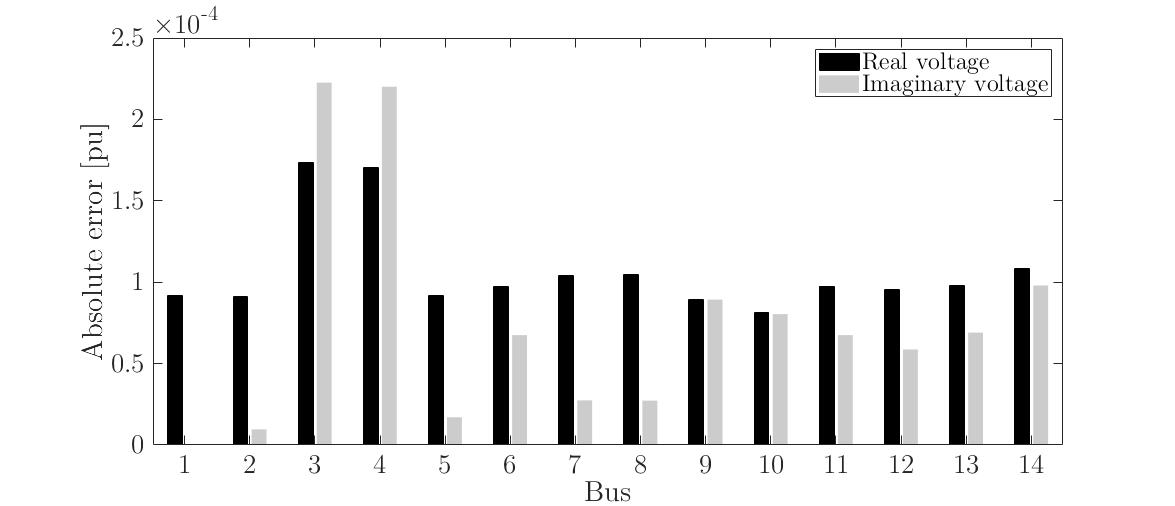}
\caption{State estimation errors for the 14 bus test case with multiple bad data}
\label{fig:multiple_bad}
\vspace{-0.5cm}
\end{figure}

One hundred simulations are performed to obtain results for different values of the measurements that have small random errors. The values of measurements with large errors are kept the same. The estimation accuracy is evaluated by calculating the values of index $\sigma_x^2$ for both test cases. These values are shown in Table \ref{tab:results_bad}. The capability of the algorithm to accurately estimate the states in the presence of bad data is clearly demonstrated, since the values of $\sigma_x^2$ are almost the same as in the case when there are no bad data. Furthermore, the absolute errors of real and imaginary estimated voltages for the case of multiple bad data, presented in Fig. \ref{fig:multiple_bad}, show very high accuracy of each individual estimated state. Finally, the average computational time, presented in Table \ref{tab:results_bad}, is still very low for both test cases, even though multiple iterations of the state estimation and bad data correction stages have to be executed. Again, this is enabled by the linearity of the proposed estimator. 

To check the scalability when bad data are present, the measurement set for the 2869 bus test case from the previous section is used, with three measurements altered by 50\%, namely the real voltage at bus 1 and active and reactive power injections at bus 455. The obtained average value of index $\sigma_x^2$ is 0.00128 and the computational time is 0.1645s. Therefore, the negative effect of bad data is suppressed and the additional computational burden imposed by running several iterations of the algorithm to correct all bad data is not significant. 
\begin{table}[t]
%\vspace{-0.4cm}
\centering
\caption{Average Performance Indices for Test Cases with Bad Data}
\label{tab:results_bad}
\begin{tabular}{ |c|c|c| }
 \hline
 \textbf{Test Case} & $\boldsymbol{\sigma^2_x}$ & \textbf{Time [s]}\\ 
 \hline
 Case 1 & 3.2167\,x\,$10^{-7}$ & 0.0004\\
 \hline
 Case 2 & 5.4783\,x\,$10^{-7}$ & 0.0008 \\ 
 \hline
\end{tabular}
\vspace{-0.3cm}
\end{table}

\section{Conclusion} \label{sec: 5}
A novel linear state estimation algorithm is proposed for the systems that comprise conventional and synchrophasor measurements. The estimator is based on the linear WLS approach, which is enabled by the adequate treatment of the RTU data and network modelling in terms of voltages and currents in rectangular form. All measurements are accounted for simultaneously and the states are estimated in rectangular coordinates. The obtained results demonstrate very high accuracy of the proposed method, as well as its low computational burden. Furthermore, utilization of the WLS framework enables the use of the LNR test that renders the algorithm bad data resilient, which comes at a low computational cost due to the linearity of the estimation stage.

\bibliographystyle{IEEEtran}

\end{document}